# Security Requirements, Counterattacks and Projects in Healthcare Applications Using WSNs - A Review

Nusrat Fatema[1], Remus Brad[1]

[1]Department of Computer Science and Electrical Engineering, Faculty of Engineering, Lucian Blaga University of Sibiu, Romania, B-dul Victoriei 10, 550024 Sibiu, Romania
knuz26@yahoo.com , remus.brad@ulbsibiu.ro

*Abstract*

*Healthcare applications are well thought-out as interesting fields for WSN where patients can be examine using wireless medical sensor networks. Inside the hospital or extensive care surroundings there is a tempting need for steady monitoring of essential body functions and support for patient mobility. Recent research cantered on patient reliable communication, mobility, and energy-efficient routing. Yet deploying new expertise in healthcare applications presents some understandable security concerns which are the important concern in the inclusive deployment of wireless patient monitoring systems. This manuscript presents a survey of the security features, its counter attacks in healthcare applications including some proposed projects which have been done recently.*

## 1. INTRODUCTION

Wireless Sensor Networks (WSNs) is the new class of infinitesimal influential computer which have been ready due to possible advances in wireless communication field. This constructs of networks which is distributed and self-organized to supervise a healthcare monitoring system [1]. This paper has come up with a survey on the security issues along with different published projects regarding the Wireless body sensor networks (WBSN).

WBSN also refers to wireless body area network (WBAN) is a wireless architecture consists of a number of body sensor units (BSUs) jointly with a single body central unit (BCU). This network consists of wearable computing devices which are under advancement. Yet these kinds of researches do not handle the argument it face while checking human body sensors.

WBAN is an on-body implanted sensor having little power included in wireless devices to facilitate secluded monitoring [2]. In real time, the situation of various patients is being monitored constantly by this architecture. To monitor this physiological situation is one of the contemporary needs to input WSN in medical system. However this kind of applications has faced different challenges while designing. For instance communication between sensors needs to be reliable and interference free and also should provide flexibility to the user. The growth in WBAN should embody the advancement of diagnosis tools of the medical monitoring scheme.

WBAN when implemented in health centres has notable advantages over the traditional user-data collection schemes. This scheme gives remedy and enhances the quality of life (Fig 1).

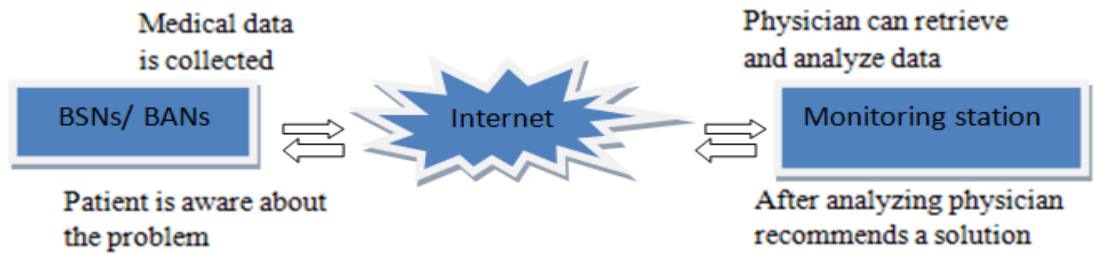

*Fig. 1* Data flow in healthcare monitoring system

## 2. PROJECTS AND RELATED WORK

Lately researcher dedicated much time to the vicinity of wireless medical care systems. Recent projects concentrates on wearable health devices. These projects are taken care of by public and private organizations which swathe many areas in healthcare like glucose level monitoring, stress monitoring, cancer detection, elderly people monitoring, ECG monitoring etc. Some recent projects are presented below:

Satire project [3], [4] (software architecture for smart attire) is a non-obtrusive wearable personal monitoring platform for data sensing, storage and upload allowing to maintain a record of patient's daily activities calculated by location and motion sensors. In [7], a patient wearing a Satire jacket records his/her daily activities. When he/ she comes in contact with a vicinity of an access mote, logged data is updated to a repository. These logged data can be used to find out the location and activity of the patient who is wearing the Satire jacket. Here security for physiological data can be a discussion for future study or work.

SMART [3] is deployed to scrutinize physiological signals of patients in the waiting areas of emergency unit. A variety of cases have been seen in waiting room where patient's health deteriorates rapidly while waiting. To solve this, this is used to collect information and wirelessly send it to a central station which accumulates and compute the data to issue an alert signal when the health deteriorates. Thus before the condition get worsens patients receives treatment.

HealthGear [3] are sensors connected to a phone by Bluetooth. It is a wearable real-time health monitoring system to analyse physiological signals.

MobiHealth/ Mobicare [3], [5], [6] is a wide area mobile healthcare scheme which allows patients to be movable while going through incessant health monitoring with the help of GPRS and UMTS cellular networks. It timely senses the patient's body data and broadcast it to client. The client aggregates the data and sends to the server for patient's improved quality of life. It make possible incessant and timely scrutinizing the physiological status of a patient. Here, MobiCare client uses the HTTP POST protocol of the application layer to send data (BSN) to the server.

CareNet [5] is deployed in an integrated wireless milieu which is used for unapproachable health care systems and it has features like high reliability, security, integration and performance.

CodeBlue [5] is a self organized platform which is easy to connect due to its adhoc architecture and it integrates different wireless devices sensor nodes into disaster response surroundings. In paper [6], a lot of body sensors like pulse oximeter, ECG sensor are connected to Zigbee-enabled transmitters individually, which corresponds with APs. Thus no intra BAN communication takes place in CodeBlue. In this approach, multiple APs are closed to a wall. Without any control from the central unit, its inter BAN communication shapes a mesh structure where physicians subscribe by multicasting to the network and the sensor devices of patients bring out all applicable information. By using these messages, physicians understands the information they need to collect. This model is flexible and secure.

The Vital Jacket [5] is a kind of wearable garment that is able to incessantly monitor Heart Rate and electrocardiogram (ECG) waves for different medical and fitness applications. Here at the same time data can be hurled via Bluetooth to PDA (Personal Digital Assistant) and is stored in a memory. Ubimon [5], [6], [7] (Ubiquitous monitoring environment for wearable and implantable sensors) aims to tackle the issues allied with wearable distributed mobile monitoring so that transient but life threatening abnormalities can be detected, captured and solved. It uses an ad-hoc network to tackle various issues.

Alarm-Net [5] presents pervasive and adaptive healthcare for nonstop scrutinizing using wireless sensors for smart healthcare by creating a history log, while preserving the patient's privacy. These sensor devices can intellect diminutive changes in decisive signals that humans might fail to notice, like heart rate, blood oxygen levels, boosting accuracy, circadian rhythm changes which may indicate changes in healthcare requirements.

AID-N [5] deals with mass casualty targeted incidents. Instead of deploying APs (Access policy) on wall, wireless repeaters are positioned to a defined emergency course. When APs flash green lights, medical staff can be aware of the accurate emergency course.

Bike Net project [8] deployed for sensing the environment and mobile bicycle activities assigned to bike area networks, while SNAP [8] architecture focuses on security. Here wireless sensors are attached to the patient. The broadcast are forwarded by a number of wireless relay nodes all over the hospital area.

eWatch [8] fits into a wrist watch provides palpable audio and visual warning while sensing and it records temperature and motion. The Tmote Sky [8] is an interesting ongoing project in the biomedical WSN. This hardware platform with integrated sensors such as persistent arterial blood pressure, ECG, epicardial accelerometer etc. As the project spotted on wireless communication and information throughput optimization, it does not tackle security issues.

## 3. SECURITY REQUIREMENTS

For suitable security mechanisms we need to comprehend the nature of applications, security requirements and its techniques properly. Then we can actually come up with an inclusive technique which can protect the system from possible safety intimidation [8]. Frequency band selection, channel modeling, antenna design and protocol design, energy-efficient hardware, QoS and reliability, real time connectivity over mixed networks, regulatory compliance, security and privacy are different kinds of issues related to WBSN. These security requirements are discussed as follows [7].

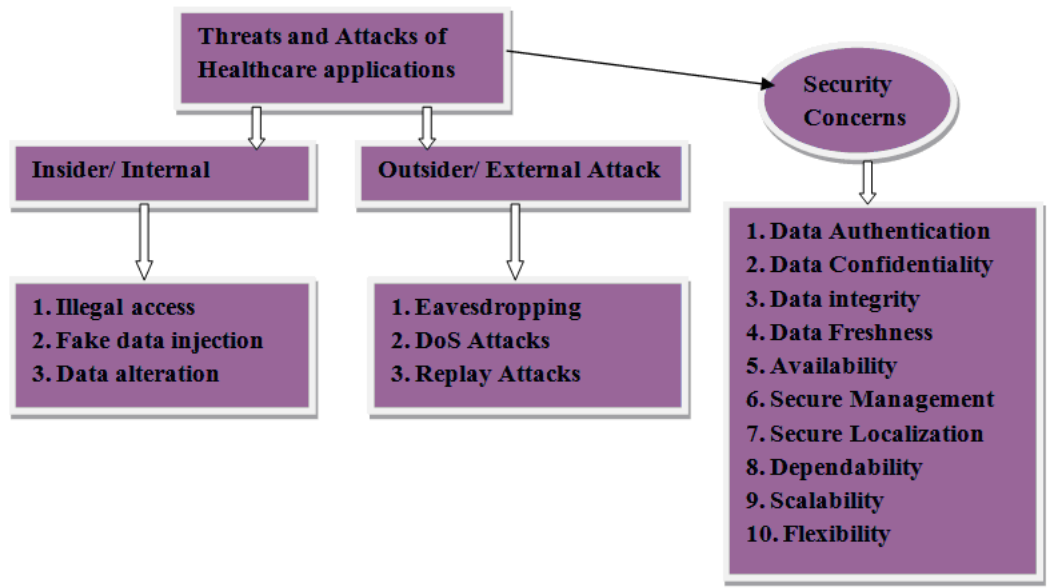

*Fig.2* *Healthcare information technology, threats, attacks and security concerns*

a) *Data Confidentiality*

It is crucial to guarantee the secrecy of messages to protect data and communication exchanges [8]. It is obligatory to guard data from disclosure. In medical applications sensitive data is send throughout the network. A foe can overhear the critical information with no trouble [10]. This over hearing can cause inexorable damage to the patient since the foe can use the obtained data for much illegal intention. As public-key cryptography encryption method is an expensive encryption method as a result the symmetric key encryption's utilization is mostly used in energy-constraint WBANs [9].

b) *Data Authentication*

It is compulsory for the sensor nodes to perceive new or replayed packets [8]. In a WBSN, it can be attained by means of symmetric techniques [10]. It substantiates the eccentricity of the original source node. A foe can change a packet stream or can modify it by amalgamating fabricated packets. The node must be an expert one to verify the details of the original source [9].

c) *Data Integrity*

When data is transmitted over an insecure WBSN the important data can be altered or moderated by a foe and this happens due to lack of integrity. This can be too dangerous in case of life-critical events. But through data authentication protocols proper integrity can be obtained further.

d) *Data Freshness*

It implies that the data is new and the frames of data are sorted and not reused. Sometimes a foe may capture data while convey and can later replay them by means of the old key to the sensor nodes. Two types of data freshness are available and they are:

- Weak freshness: assures fractional data frames ordering but does not assure delay.

- Strong freshness: guarantees both delay and data frames ordering.

*e) Availability*

It ensures patient's information to be obtainable to the physician for all time. A foe may target the availability of a WBSN by capturing or turning off an ECG node, which may cause death to a patient [10]. Whereas designing security mechanisms that address the above requirements we need to know the following factors which differentiate WBSN from other types of sensor networks [8]:

- It is necessary to support multiple users in different roles having various privacy interests and decision making power.
- It is necessary for the security protocol to add a small communication overhead because throughput is decisive for those networks.
- Mobility of the patient must be maintained so that security mechanisms can be adapted to dynamic topologies.

*f) Secure Management*

It is required because it gives key distribution to the nodes to allow both encryption and decryption operation.

*g) Secure Localization*

Precise estimation of the patient's location is needed in most WBSN applications. Due to lack of smart tracking mechanisms permits a foe to send erroneous reports concerning the patient's location either by reporting fake signal strengths or by using rerun the signals [9].

*h) Dependability*

In most medical cases, unable to retrieve precise data is a life-threatening issue. To tackle threats caused by the network dynamics, fault tolerance is requisite that is having patient's data willingly retrievable even under failure of nodes or malicious modifications [11].

*i) Scalability*

Distributed access control method must be scalable with the increase number of users in following ways:

- To have low management overhead of the APs, which shall be set up and modified
- To have low computation and storage space overhead [11].

*j) Flexibility*

A basic obligation is that patient should have the flexibility to designate APs to control the medical data within WBAN. For example, on-demand authorization to interpret patient's data can be given for the time being to an available physician who is not on the permissible list when an emergency occurs. Inability or irresponsiveness in becoming accustomed the access rules may threaten a patient's life [11].

## 4. THREATS AND ATTACKS

Security breach occurs mostly in sensor networks of healthcare applications. These specific breaches that a biomedical WBSN has to face can be labeled as outsider and insider attacks [8].

### 4.1 Insider/ Internal attack

From a security point, it is more unsafe where foes captures a node and reads its reminiscence and get hold of its key material and falsify node messages. A foe can commence several kinds of attacks by knowing the legal keys:

- Illegal access to health data: If without valid authentication any important patient's data is accessed then it might cause problems such as damaging significant data
- Fake data injection: where the intruder insert fake results which are very much different from the realistic health data persistent by the sensors.
- Careful reporting: where an enemy freezes the report of actions by throwing authenticate packets which pass through the node.
- Data alteration: where the foe alter data of a patient which leads to erroneous diagnosis and treatment.

### 4.2 External attack (intruder node attack)

The principle purpose of attacks is to filch valuable personal data. Once the foe is aware of the personal health data they try to steal it. Authentication and encryption practices can prevent such an invader to gain any access to WBSN. The attacks are as follows:

- Passive eavesdropping: While routing the data packets it can take place. The foe may alter the destination of packets or can make routing conflicting [3]. Here, the foe can also steal health data by snooping to the wireless media.
- Denial of service attacks: where a foe attempts to upset the operation of the network by broadcasting lofty energy signals or by jamming the communication media between nodes.
- Replay attacks: where the foe captures messages switch over between legitimate nodes and replays them to modify the aggregation results.

The authors in [3] have mentioned in details few types of attacks in health monitoring that is modification, forging of alarms on medical data, overhearing, denial of service, location tracking and activity tracking of users, physical tampering with devices and jamming attacks.

## 5. COUNTER ATTACKS AND DETECTION SYSTEMS

As mentioned in the research paper [3, 8], by ensuring only genuine devices can generate and insert data to the scheme and by averting illegal modifications of data can evade many of the previously argued attacks. To counter these threats the following security measures can be applied.

### 5.1 Encryption & Authentication

Due to the perceptive nature of medical application we should be careful with encryption mechanism and constant monitoring of the network. Both offer guards for mote-class outsider threats. Spoofing attacks can be barred by using superior encryption techniques and suitable authentication for communication Cryptography is not sufficient for preventing the insider attacks. It is an open quandary for further research. The security safeguard measures are practical in three stages – administrative, physical and technical.

- Administrative level security: A well defined hierarchy having tough authentication measures may avert security breaches. Hence these security measures must comprise of various types of access means so that only certified users have the permission to access in the information.

- Physical level security: Devices at this level may be susceptible to stealing and tampering hence vigilant designing is compulsory to formulate them tamper proof. Only authorized public should be permitted to physically handle the devices while in operation can be another preventive measure.

- Technical level security: Here at this level security checks are needed for wireless means and also for proliferation of data. More influential and dominant motes should be designed to support the escalating requirements for computation and communication [7].

### 5.2 Securing routing data

Secure routing is needed as a security measure when the data is forwarded to some remote host (physician or some hospital computers). Foes can cause routing discrepancy resulting in wrong destinations and getting of incorrect data. Therefore correct routing protocol and management is indispensable to stay away from such attacks. Watchdog helps to supervise the route properly and provides secure network. Validation technique is used to prevent this routing attack.

### 5.3 Intrusion Detection

As wireless networks are susceptible to intrusion, detection and prevention techniques are hence a must. An automated method is established that identifies the source of an attack and engenders an alarm to warn administrator so that suitable defensive actions can take place. But intrusion detection for sensor networks requires considering constraints forced by the limited resources of nodes. Still it is a major solution for biomedical sensor networks [8].

## 6. REQUIREMENTS FOR WIRELESS MEDICAL SENSORS

The main requirements for wireless medical sensor networks are:

a) *Reliability and Robustness*

For medical diagnosis and treatment sensors must function with sufficient reliability and must be vigorous to yield high-confidence data [12].

b) *Wearability*

To attain non-invasive health monitoring, WBSN should be lightweight and small. The weight and size of sensors is determined by the mass and size of batteries. But again at the capacity of a battery is interrelated to its dimension. We can anticipate that further technology advances in miniaturization of integrated circuits will assist designers to expand wearable sensors.

*c) Interoperability*

The wireless medical system must give middleware interoperability among devices to carry out unique relations among devices, and to collect robust WBSN depending on the user's state of health [3]. Due to the heterogeneity of the system, the communication between devices may use numerous bands and utilize different protocols. For instance, motes may use ISM or unlicensed bands for general telemetry. Implanted medical sensors are using allocated licensed band in order to avoid interference [12].

*d) Real-time data acquisition and analysis*

Efficient communication, data acquisition and examination are essential. Event ordering, time-stamping, synchronization, and rapid response in emergency circumstances will all be requisite.

*e) Reliable communication*

The communication necessities of diverse healthcare sensors vary with requisite sampling rates. An approach is to advance reliability is to go further than telemetry by performing on-sensor signal processing. For instance without transferring raw data from an ECG sensor, we can perform feature extraction on sensor and move information about an event. So a vigilant trade-off between computation and communication is essential for finest system design.

*f) New node architectures*

The amalgamation of diverse sensors, RFID tags and back channel extended haul networks require new and modular node architectures.

## 7. CONCLUSION AND FUTURE WORK

WBSN is one of the significant elements of the health care system and a promising technology. After analyzing the whole manuscript we have come to a verdict that the counter attack which has been researched till now is not enough and it is an unbolt challenge still needed further analysis. Existing medical applications based on WSNs are the research scheme with first-rate potential for utilization that is the future of WSNs and their medical appliances looks exceedingly promising. Security issues are a significant area, and there still remain a number of extensive challenges to overcome. This appraisal paper will motivate other researcher to defeat those challenges with improving the quality of service (QoS), security, privacy, reliability, fault tolerance and interoperability.

*Authors*

**Nusrat Fatema**   is the student of Lucian Blaga University of Sibiu, Romania, studying MSc in Embedded Systems.

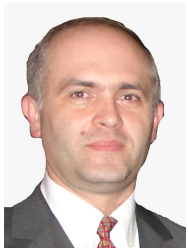

**Remus Brad** received an Engineer Diploma degree in Automation and Computer Science from the Lucian Blaga University of Sibiu, Romania, a M.S. degree from Université Pierre et Marie Curie, Paris France in Artificial Intelligence and a PhD. from Technical University of Cluj-Napoca Romania. Since 1994 he has joined the Department of Computer Science at the Lucian Blaga University of Sibiu, Romania. His current research interests include image processing, motion estimation and biomedical imaging. Dr. Remus Brad is a member of the IEEE Signal Processing Society.